# Modelling proton-induced reactions at low energies in the MARS15 code[*]

**Igor L. Rakhno[1], Nikolai V. Mokhov[1], Konstantin K. Gudima[2]**

[1]Fermi National Accelerator Laboratory, Batavia, Illinois 60510, USA
[2]Institute of Applied Physics, National Academy of Sciences, Cisineu, Moldova

**Abstract**

*An implementation of both the ALICE code and TENDL evaluated nuclear data library in order to describe nuclear reactions by low-energy projectiles in the Monte Carlo code MARS15 is presented. Comparisons between results of modelling and experimental data on reaction cross sections and secondary particle distributions are shown as well.*

______________________________________________________________________

**Introduction**

Correct prediction of secondary particles, both neutral and charged ones, generated in proton-nucleus interactions below a few tens of MeV is required for various applications. The latter include, *e.g.*, radiation studies for front-end of many proton accelerators, energy deposition studies for detectors, radiation damage calculations, *etc*. Cascade models of various flavours fail to properly describe this energy region (see Figure 1). Therefore, we opted to use the TENDL library developed by the Nuclear Research and Consultancy Group [1]. The evaluated data is provided in the ENDF/B format in the energy range from 1 to 200 MeV, and the library is regularly updated since 2008. In addition, a much more time-consuming approach utilized in a modified code ALICE [2] was also looked at. For both the options, the energy and angle distributions of all secondary particles are described with the Kalbach-Mann systematics. The following secondaries are taken into account: gammas, neutrons, protons, deuterons, tritons, $^3$He and $^4$He. The energy and angular distributions of all generated residual nuclei—including unstable ones—are accounted for as well.

Various comparisons with experimental data for both the options are presented. The corresponding processing and modeling software was written in C++ which provides substantial flexibility with respect to the computer memory used. In addition, the initialization of required evaluated data is performed dynamically whenever the modeling code encounters a nuclide not accounted for yet. The latter feature enables us to significantly reduce the amount of requested memory for extended systems with large number of materials.

**Figure 1: Calculated and measured [3] neutron production cross section for aluminum at low incident proton energies**

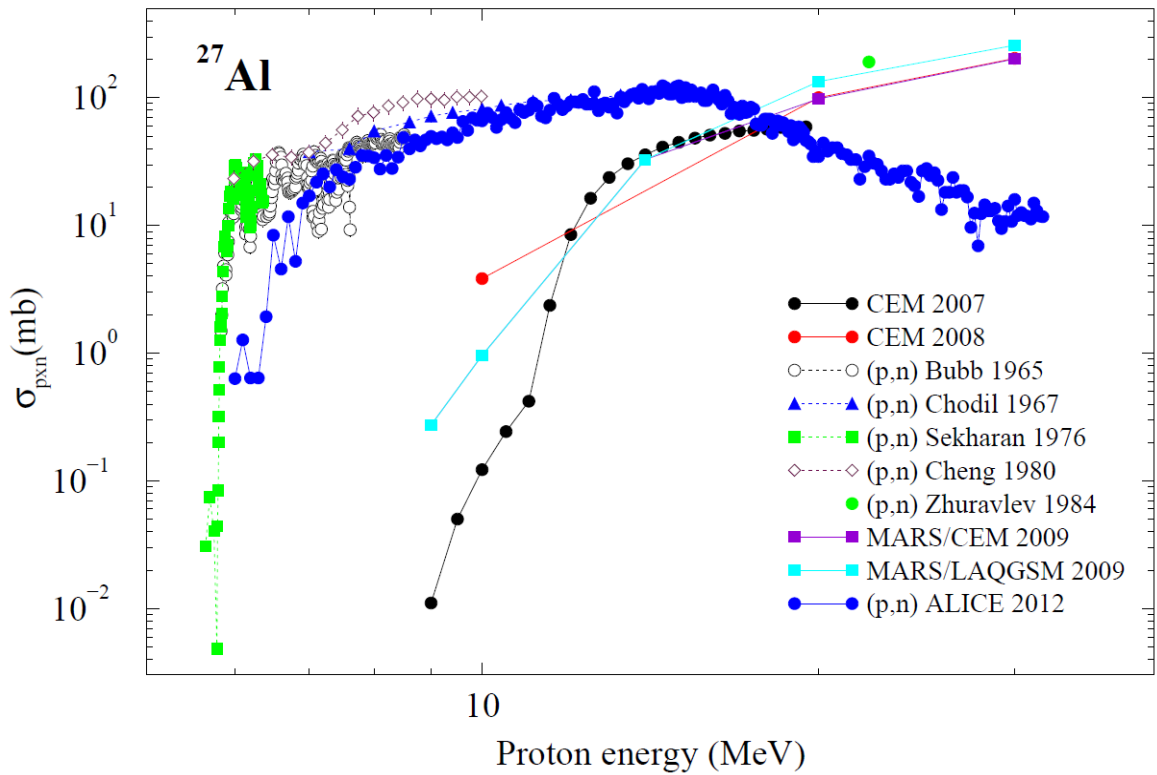

**Details of the implementation and formalism**

*TENDL library*

The TALYS-based evaluated nuclear data library (TENDL) contains data for direct use in both basic physics and applications, and it is updated annually starting from 2008. The evaluations were performed for practically entire periodic table except for hydrogen and helium. Projectile kinetic energies are from 1 to 200 MeV. The library contains data for both stable and unstable target nuclei—all isotopes which live longer than 1 second were taken into account. At present, the list includes about 2400 isotopes.

In fact, the TENDL library contains data not only for protons as projectiles, but also for light ions (d, t, $^3$He, $^4$He), neutrons and gammas. In current implementation, the data for protons and light ions are used when modelling low-energy reactions in MARS15 code [4-5]. It should be noted also that at present only inelastic collisions are taken into account. In other words, for every interaction the following information is extracted from the library: (i) total inelastic cross section; (ii) energy and angular distributions of all above mentioned secondary particles and residual nuclei ; (iii) yields of all the secondaries.

The TENDL library is a collection of files in both ENDF/B and ACE format. We opted to use the source—ENDF/B format—because it is more human friendly which is a very important feature at the development and debugging stages. At the same time, there is little difference between the two formats with respect to storage requirements.

*ALICE code*

An alternative approach—using an event generator—was explored as well. For this purpose, the nuclear model code ALICE [2] based on a hybrid model of pre-compound decay, Weisskopf-Ewing evaporation and Bohr-Wheeler fission models was employed. It was re-designed in order to be used as an event generator for nucleon, photon and heavy-ion nuclear reactions at incident energies from 1 to 20-30 MeV matching CEM and LAQGSM at energies above 20-30 MeV in MARS15. At present, this option looks much more time consuming, but potentially it can offer some advantages for applications where accuracy of the full exclusive modeling is of major importance, *e.g.* when modelling a detector performance.

*Kalbach-Mann systematics*

Nowadays, continuum energy-angle distributions of secondary particles generated in low energy interactions are described mostly with the Kalbach-Mann systematics [6] represented by Equation (1).

$$f(\mu_b, E_a, E_b) = 0.5 * f_0(E_a, E_b) \left[ \frac{a}{\sinh(a)} [\cosh(a\mu_b) + r(E_a, E_b)\sinh(a\mu_b)] \right], \quad (1)$$

where $a = a(E_a, E_b)$ is a parametrized function, $r(E_a, E_b)$ is the pre-compound fraction as given by the evaluator, $\mu_b$ is cosine of scattering angle, $E_a$ and $E_b$ are energies of incident projectile and emitted particle, respectively, and $f_0(E_a, E_b)$ is total probability of scattering from $E_a$ to $E_b$ integrated over all angles. The function $a(E_a, E_b)$ depends mostly on the emission energy, and there is also a slight dependence on particle type and the incident energy at higher values of $E_a$.

## Comparisons between experimental data and results of modelling

### Reaction cross sections

Comparisons between calculated and measured reaction cross sections for several light and medium target nuclei are shown for incoming protons in Figures 2-5 and for incoming deuterons in Figure 6.

**Figure 2: Calculated (ALICE, TENDL) and measured [3] single neutron production cross section on $^{65}$Cu**

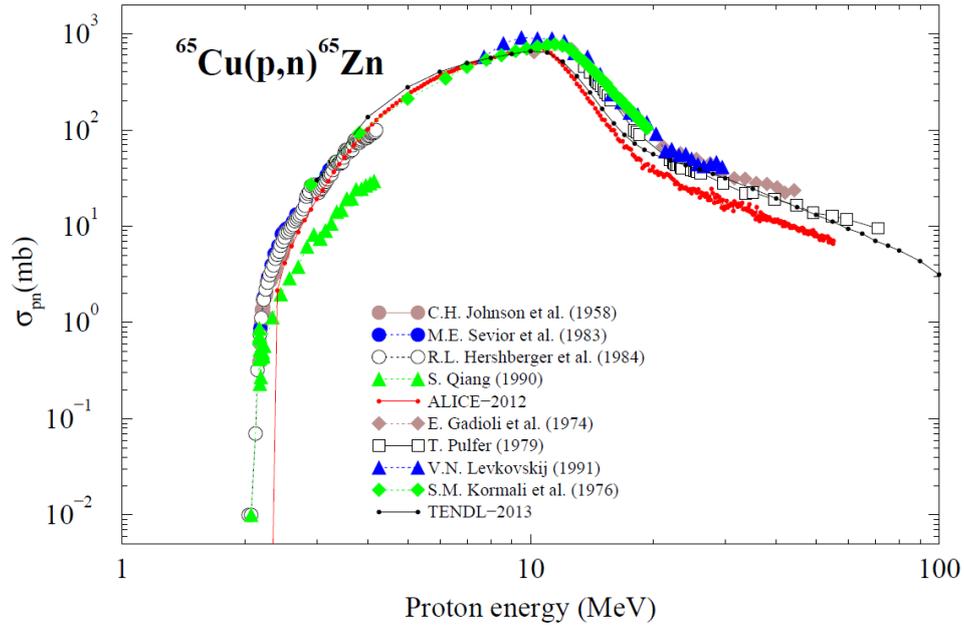

**Figure 3: Calculated (ALICE, TENDL) and measured [3] neutron production cross section on $^{65}$Cu**

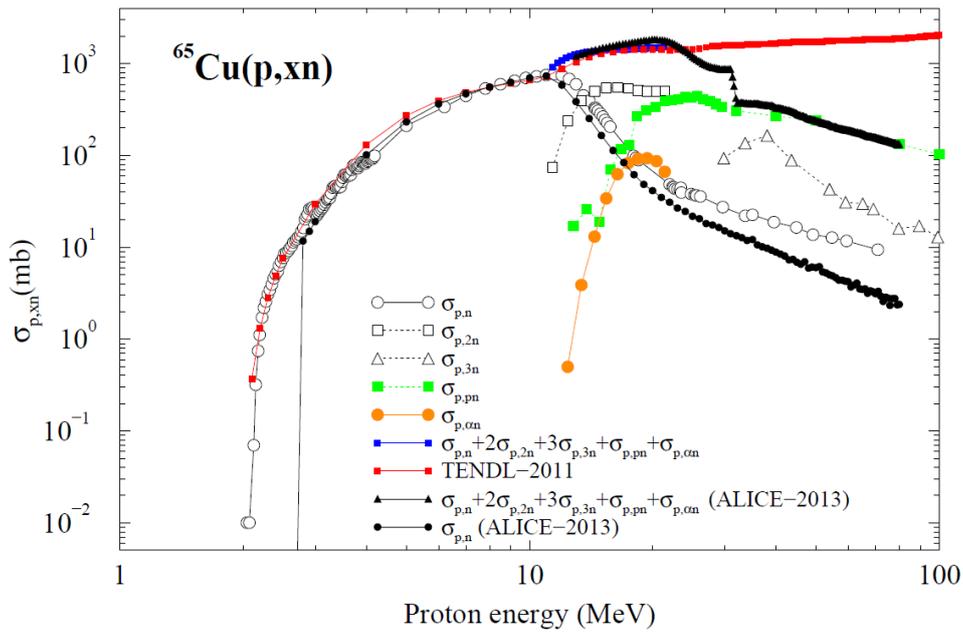

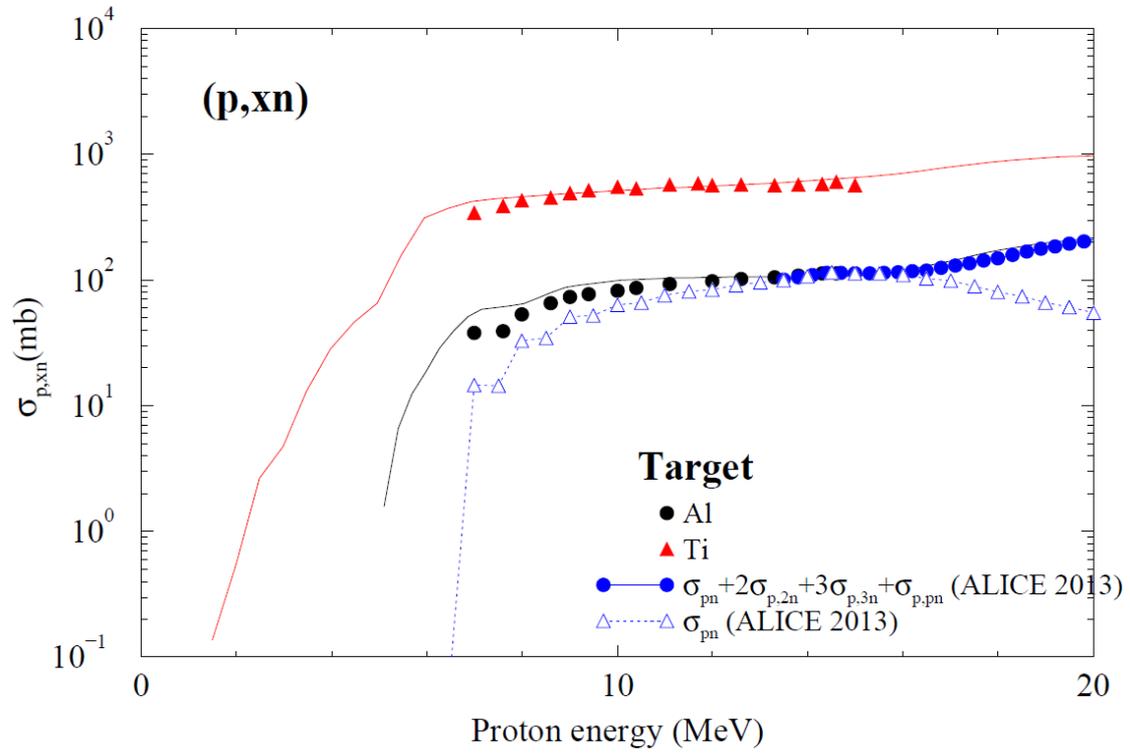

Figure 4: Calculated (red and black lines - TENDL) and measured [3] neutron production cross section on Al and Ti nuclear targets

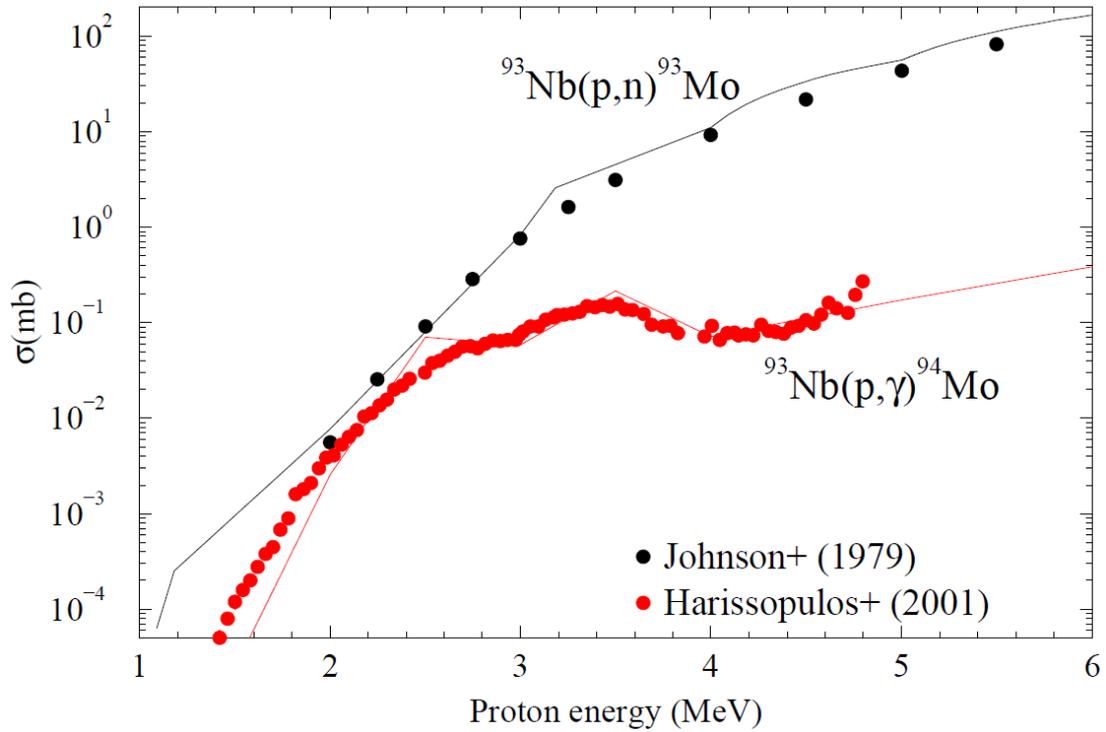

Figure 5: Calculated (lines - TENDL) and measured [7-8] reaction cross sections on $^{93}$Nb

**Figure 6: Calculated (lines - TENDL) and measured (circles [9] and triangles [10]) deuteron-induced reaction cross sections on $^{93}$Nb**

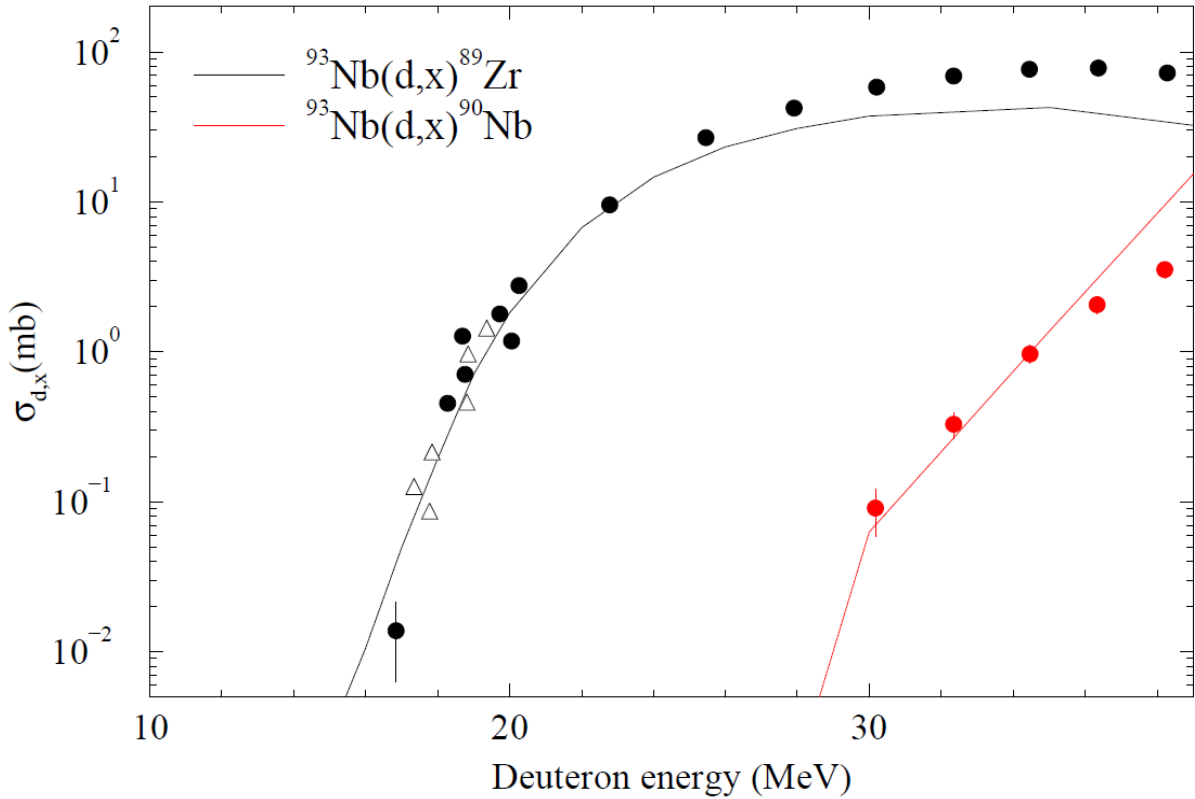

### Energy distributions of secondary particles

Energy spectra of secondary neutrons are shown in Figures 7-9. One can see in Figure 7 that at low secondary energies—where the major neutron emission takes place—the TENDL library provides pretty reliable predictions for energy distributions. At the same time, at the very low projectiles energies—about 10 MeV and lower—the library does not reproduce the significant nuclear structure effects observed in the experimental data (see Figure 9). However, for high-energy applications the accuracy provided by TENDL library is quite adequate.

### Angular distributions of secondary particles

Angular distributions of secondary neutrons are shown in Figures 10 and 11. One can see that both the library and ALICE-based generator can reproduce well the general features of the angular distributions: close to isotropic distributions at low emitted neutron energies and forward-peaked fistributions at higher secondary energies. The observed numerical agreement between calculations and experimental data in these two cases of medium and heavy nuclei is also pretty good.

### Transition region

In order to provide smooth transition between the low-energy region described, *e.g.*, with the TENDL evaluated data library and higher energies where cascade models work well, several cases were studied (see Figures 12 thru 14).

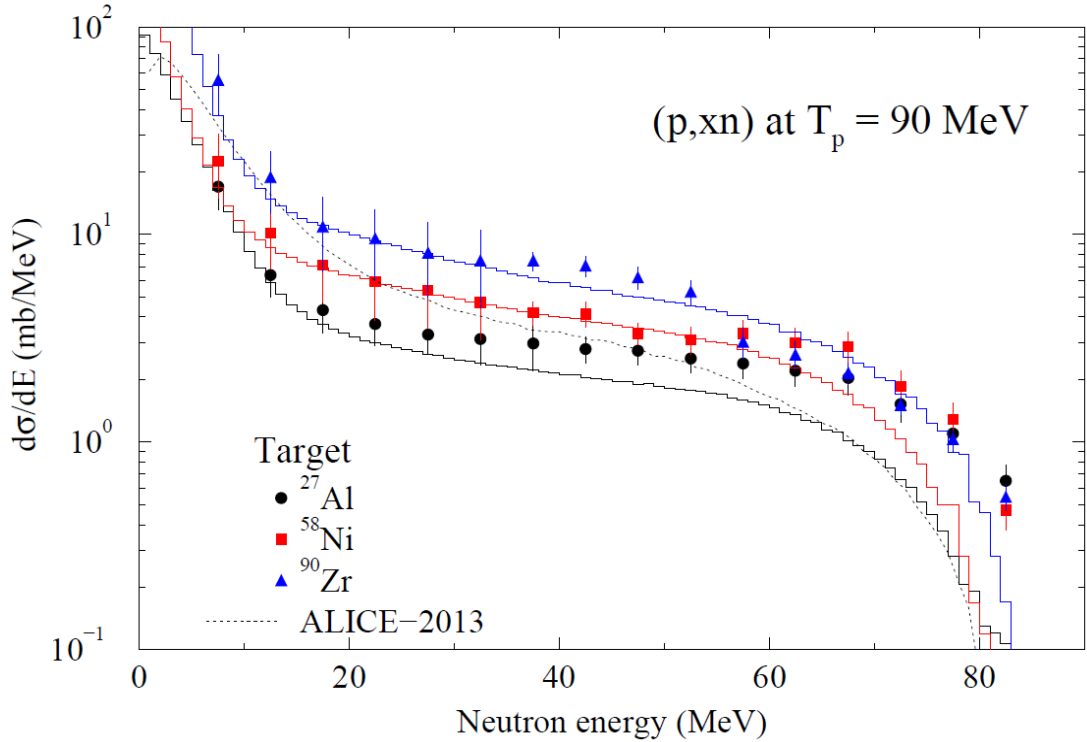

**Figure 7:** Calculated (solid lines – TENDL) and measured [3] energy distributions of secondary neutrons

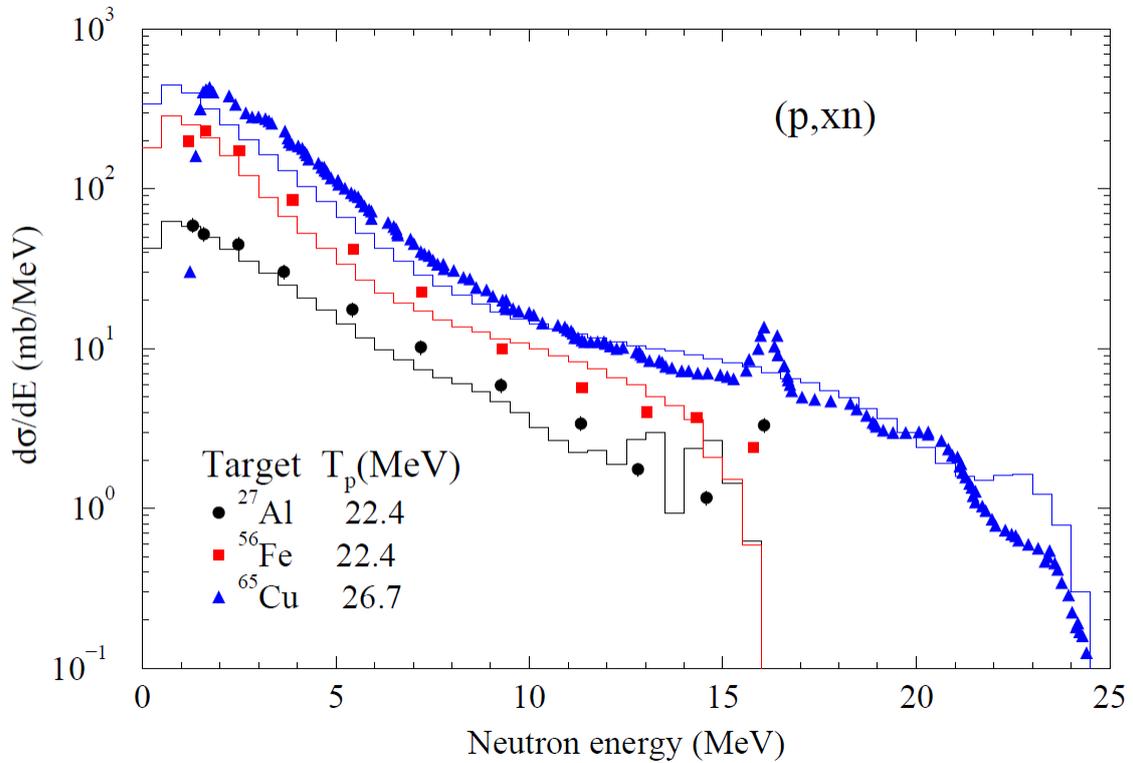

**Figure 8:** Calculated (lines – TENDL) and measured [3] energy distributions of secondary neutrons

**Figure 9:** Calculated (lines – TENDL) and measured [3] energy distributions of secondary neutrons

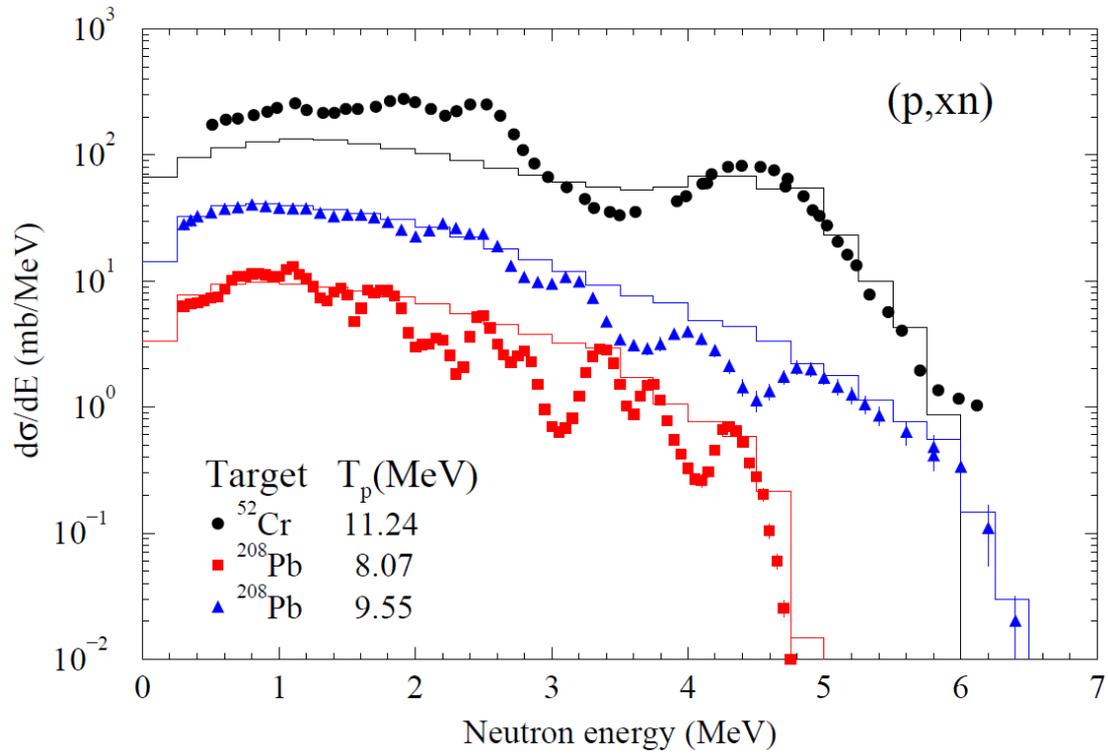

**Figure 10:** Calculated (solid lines – TENDL) and measured [3] angular distributions of secondary neutrons from $^{65}$Cu

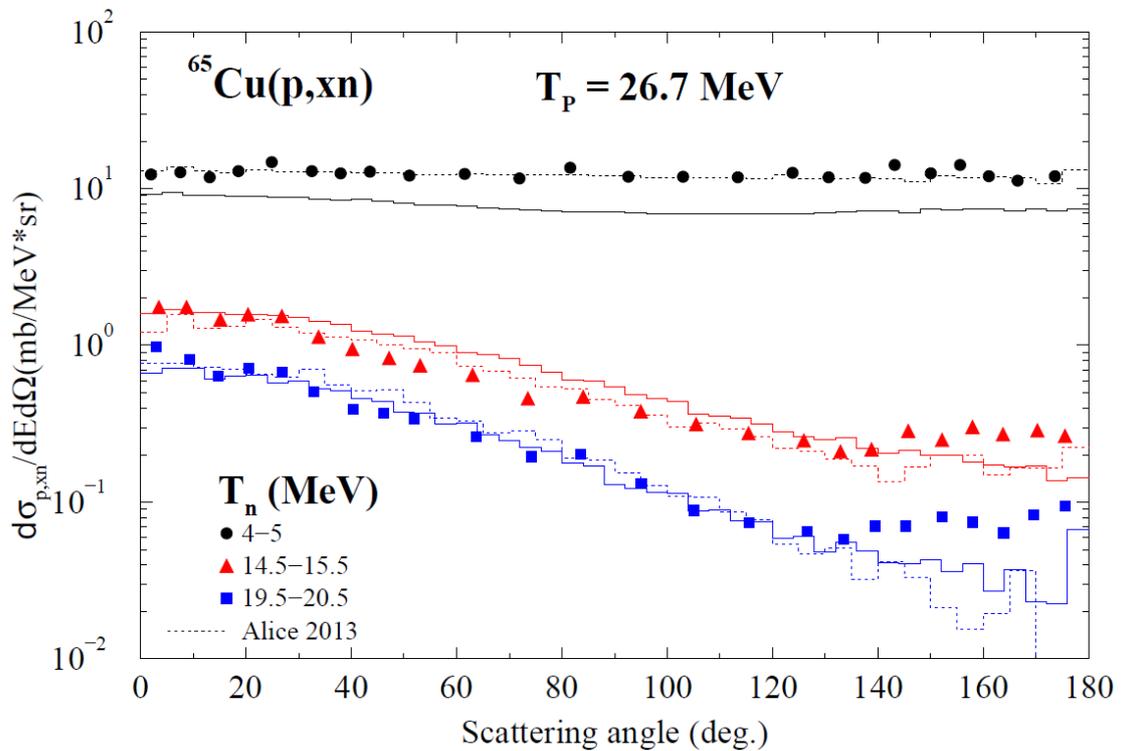

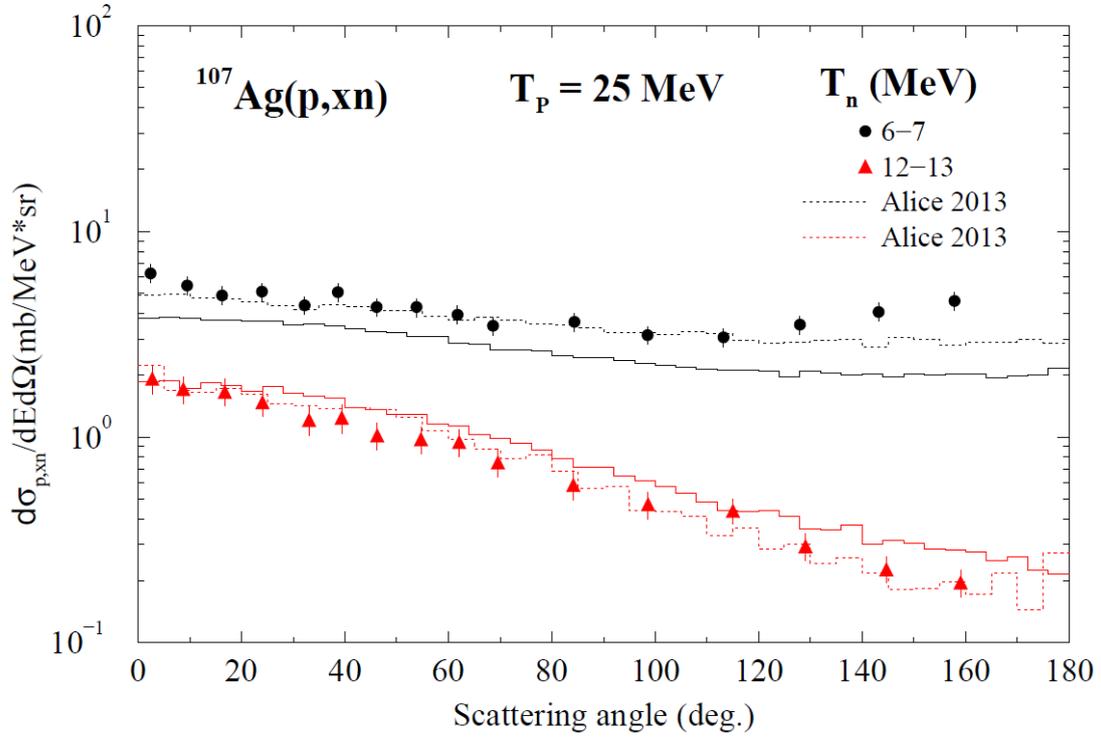

**Figure 11:** Calculated (solid lines – TENDL) and measured [3] angular distributions of secondary neutrons for $^{107}$Ag

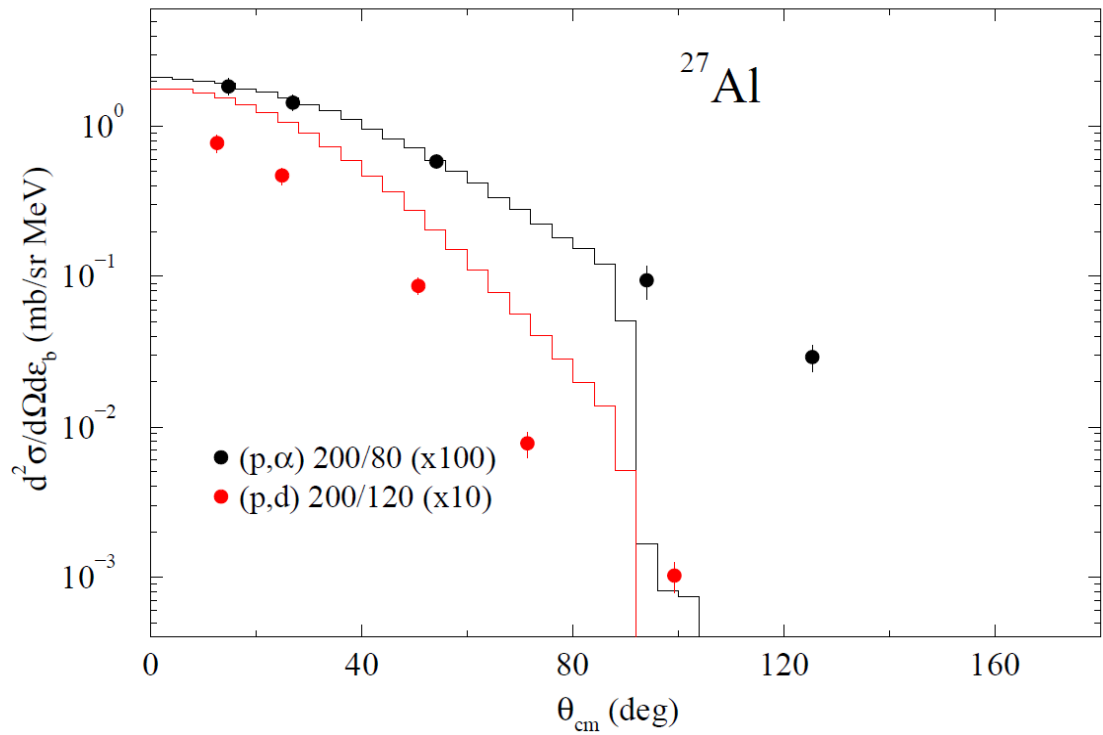

**Figure 12:** Calculated (lines – TENDL) and measured [6] angular distributions of secondary deuterons and α-particles from $^{27}$Al (projectile/ejectile energies are given in MeV)

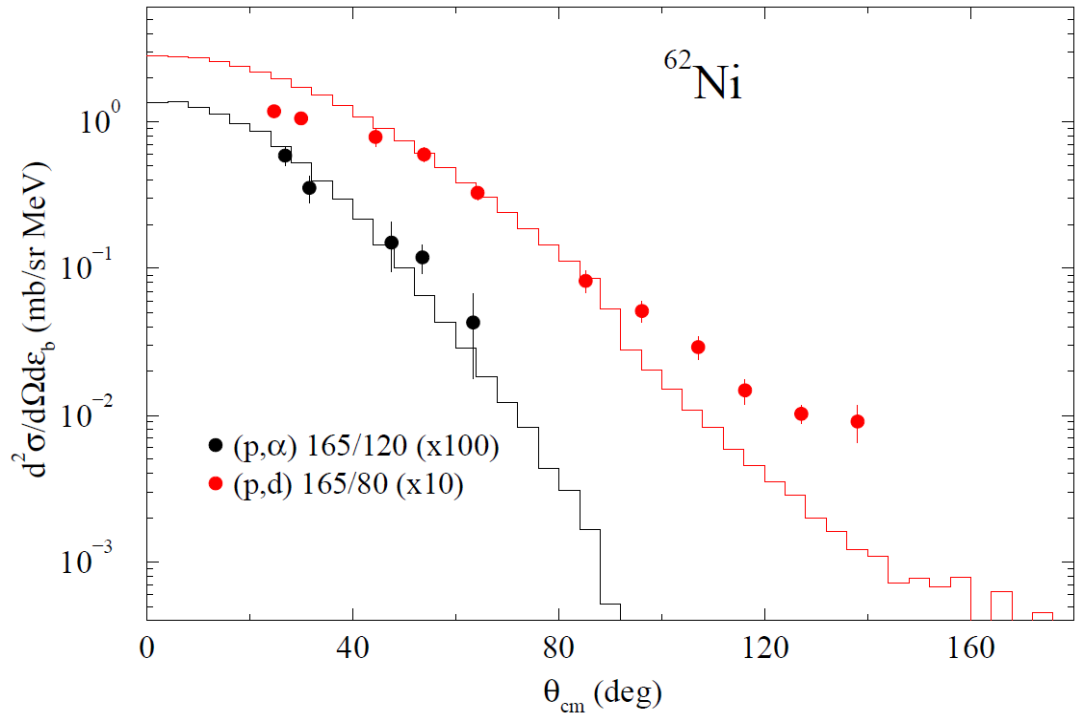

**Figure 13: Calculated (lines – TENDL) and measured [6] angular distributions of secondary deuterons and α-particles from $^{62}$Ni (projectile/ejectile energies are given in MeV)**

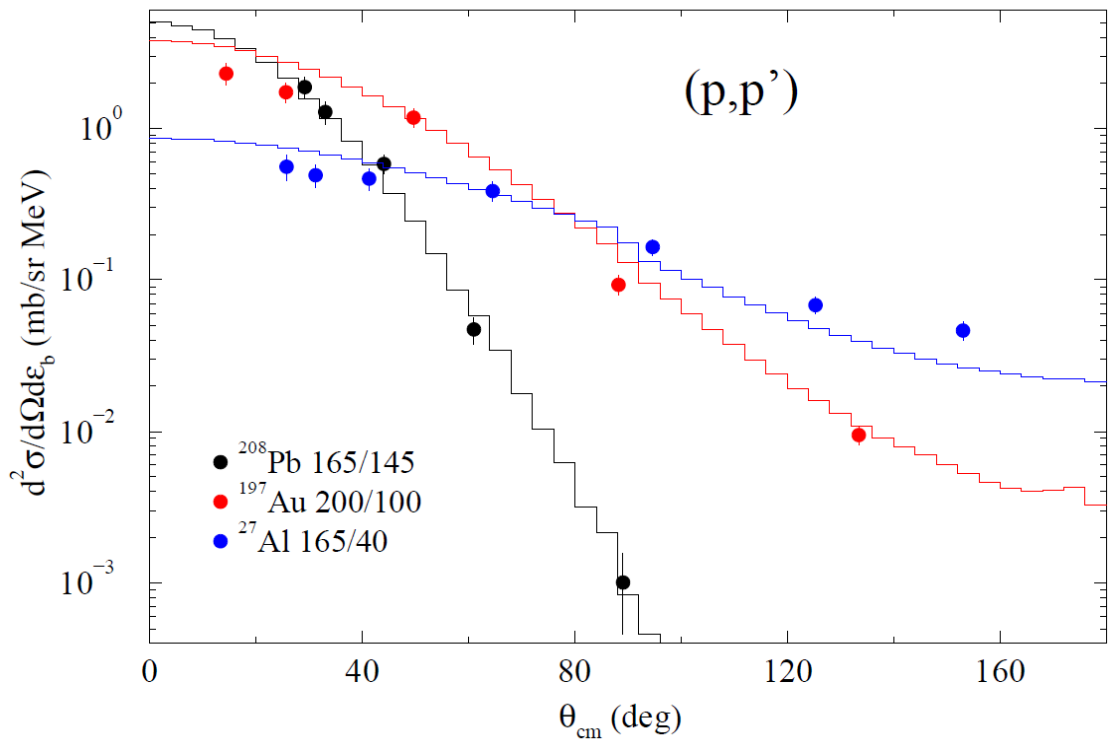

**Figure 14: Calculated (lines – TENDL) and measured [6] angular distributions of secondary protons for light, medium and heavy target nuclei (projectile/ejectile energies are given in MeV)**

One can see that the transition region between low energies described with TENDL library and higher energies can be very broad because the library provides a good description of experimental data up to projectile energy of 200 MeV. At the same time, the transition region is expected to be dependent on target mass number because, in general, quality of the Kalbach-Mann systematics gets worse with decreasing target mass number (see Figure 12). The work on proper description of the transition region between low and higher energies is currently in progress.

### Acknowledgements

This work was supported by Fermi Research Alliance, LLC, under contract No. DE-AC02-07CH11359 with the U.S. Department of Energy.